\documentclass[preprint,authoryear]{elsarticle}

\usepackage[colorlinks]{hyperref}
\usepackage{graphicx}
\usepackage{amsmath}
\usepackage{hhline}
\usepackage{booktabs}   
\usepackage{multirow}
\usepackage{xcolor}
\usepackage{xurl}
\usepackage{subcaption}
\usepackage{float}
\usepackage{threeparttable}
\usepackage{colortbl}
\usepackage{ulem}  

\usepackage{multirow}
\definecolor{gray}{gray}{0.9}

\newcommand{\specialcell}[2][c]{%
  \begin{tabular}[#1]{@{}c@{}}#2\end{tabular}}

\journal{Medical Image Analysis}

\biboptions{authoryear}

\begin{document}

\begin{frontmatter}

\title{Interpretable Vertebral Fracture Quantification via Anchor-Free Landmarks Localization}



\author[ira,skoltech]{Alexey Zakharov\fnref{equal}}
\author[ira,iitp]{Maxim Pisov\fnref{equal}}
\author[ira]{Alim Bukharaev\fnref{equal}}
\fntext[equal]{Equal contribution}
\author[radiology]{Alexey Petraikin}
\author[indep]{Sergey Morozov}
\author[airi]{Victor Gombolevskiy}
\author[ira,skoltech]{Mikhail Belyaev\corref{mycorrespondingauthor}}
\cortext[mycorrespondingauthor]{Corresponding author}
\ead{m.belyaev@skoltech.ru}

\address[ira]{IRA Labs Ltd, Moscow, Russia}
\address[skoltech]{Skolkovo Institute of Science and Technology, Moscow, Russia}
\address[iitp]{Kharkevich Institute for Information Transmission Problems, Moscow, Russia}
\address[radiology]{Research and Practical Clinical Center for Diagnostics and Telemedicine Technologies of the Moscow Health Care Department}
\address[indep]{Osimis SA, Liège, Belgium}
\address[airi]{Artificial Intelligence Research Institute, Moscow, Russia}

\begin{abstract}
Vertebral body compression fractures are early signs of osteoporosis. Though these fractures are visible on Computed Tomography (CT) images, they are frequently missed by radiologists in clinical settings. Prior research on automatic methods of vertebral fracture classification proves its reliable quality; however, existing methods provide hard-to-interpret outputs and sometimes fail to process cases with severe abnormalities such as highly pathological vertebrae or scoliosis. 
We propose a new two-step algorithm to localize the vertebral column in 3D CT images and then detect individual vertebrae and quantify fractures in 2D simultaneously. We train neural networks for both steps using a simple 6-keypoints based annotation scheme, which corresponds precisely to the current clinical recommendation. Our algorithm has no exclusion criteria, processes 3D CT in $2$ seconds on a single GPU, and provides an interpretable and verifiable output. The method approaches expert-level performance and demonstrates state-of-the-art results in vertebrae 3D localization (the average error is 
$1$ mm), vertebrae 2D detection (precision and recall are $0.99$), and fracture identification (ROC AUC at the patient level is up to $0.96$). 
Our anchor-free vertebra detection network shows excellent generalizability on a new domain by achieving ROC AUC $0.95$, sensitivity $0.85$, specificity $0.9$ on a challenging VerSe dataset with many unseen vertebra types.
\end{abstract}

\end{frontmatter}


\section{Introduction}    

Osteoporosis is a systemic skeletal disease manifested by low bone mass and deterioration of bone microarchitecture followed by increased bone fragility. The clinical manifestation of osteoporosis is the occurrence of bone fractures \citep{kanis2019european} which are common in older adults and resulted in more than two million Disability Adjusted Life Years in Europe \citep{johnell2006estimate}. Typically, osteoporotic fractures are localized in the spine, hip, distal forearm, and proximal humerus. 

Osteoporotic fracture risk models are becoming increasingly popular, while bone mineral density (BMD) is a major contributing factor. However, the prevalence and severity of vertebral compression fractures (VCFs) are predictive for the risk of new osteoporotic fractures independently of bone mineral density (BMD) measurements \citep{malgo2017value}.  
In particular, vertebrae fractures usually occur before hip fractures \citep{riggs1995worldwide} and dramatically increase the probability of the subsequent fractures \citep{klotzbuecher2000patients}; thus can be used as an early marker of osteoporosis.  

Medical imaging, such as Computed Tomography (CT), is a useful tool to identify VCFs \citep{lenchik2004diagnosis}, especially as an incidental finding. However, radiologists usually analyze CT by navigating through axial slices as, first, computed tomography produces axial slices by design, and second, this view is sufficient to analyze the majority of pathological conditions, e.g., lung diseases. In contrast, vertebral fractures identification is an exception and must be analyzed in the sagittal plane. Multiplanar reconstructions are not generated automatically in many hospitals and require some additional manual steps from radiologists \citep{gossner2010missed}. As a result, radiologists frequently miss fractures, especially if they are not specializing in musculoskeletal imaging, with the average error rate being higher than 50\% \citep{mitchell2017reporting}. At the same time, rapidly evolving lung cancer screening programs or active usage of CT for COVID-19 diagnosis and management provide a solid basis for opportunistic screening of vertebral fractures.

The medical image computing community thoroughly investigated fractures detection and/or classification on vertebrae-level  \citep{roth2016deep,valentinitsch2019opportunistic,burns2017vertebral,antonio2018vertebra,grading-loss}, whole study-level \citep{tomita-rnn,zebra,seq-to-seq-backbone}, or jointly on both levels \citep{nicolaes-segmentation,yilmaz2021automated}, see Section \ref{sec:previous} for more details.
Many of these approaches require prior vertebrae detection \citep{antonio2018vertebra,valentinitsch2019opportunistic,nicolaes-segmentation,grading-loss}, or spine segmentation \citep{burns2017vertebral,roth2016deep,zebra}. Though both problems are active areas of research with prominent results, fractured vertebrae are the most complex cases for these algorithms \citep{sekuboyina2017attention}, and even good average detection/segmentation accuracy may not be sufficient for accurate fracture estimation. As a result, researchers had to exclude some studies from the subsequent fracture classification due to errors in prior segmentation \citep{valentinitsch2019opportunistic}, 
or due to scoliosis \citep{tomita-rnn}. 

\begin{figure}[t!]
    \begin{center}
      \includegraphics[width=\linewidth]{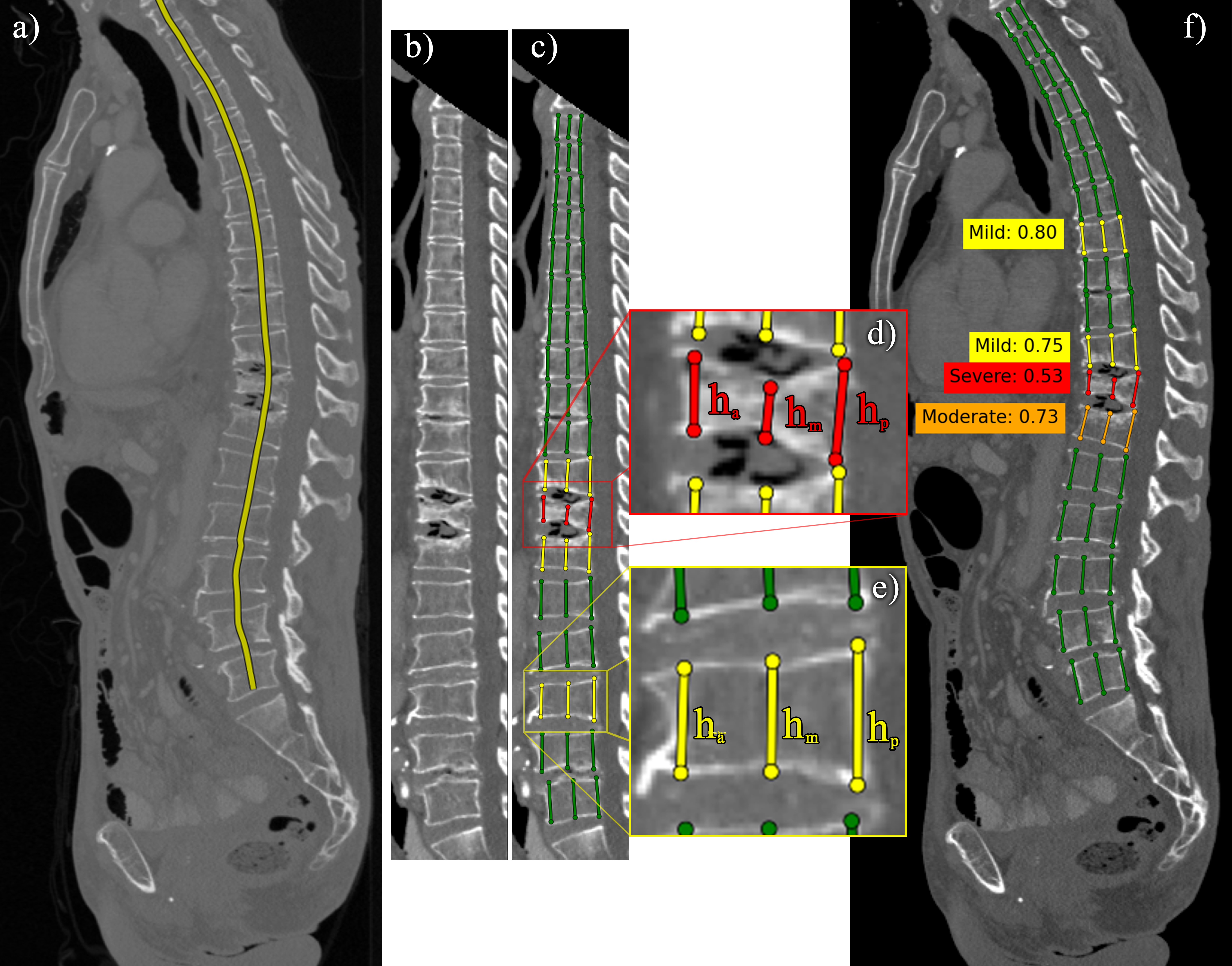}
      \caption{
      Overview of the proposed model.
      \textbf{Step 1}: a) localizing vertebrae centers in 3D CT (a sagittal projection is shown); 
      b) generating a new 2D image via spine `straightening'.
      \textbf{Step 2}: c) identifing key-points and the corresponding heights;
      d-e) a closer look at some vertebrae (colors denote the fracture severity).
      \textbf{Finally}: f) the original image with vertebrae types (right side) and their estimated fracture severities (left side).
       }
      \label{fig:method}
    \end{center}
\end{figure}

The second important issue is the mismatch between computer science problem statements and the radiological way to define fractures. The Genant scale \citep{genant} is a widely used medical criterion 
recommended by the International Osteoporosis Foundation \citep{iof_recs}. It relies on the measurements of  $h_a, h_m, h_p$ - the anterior, middle and posterior heights of vertebral bodies (Fig. \ref{fig:method}d, \ref{fig:method}e):
\begin{equation}
    \label{eq:genant}
    G = \frac{\min \{h_a, h_m, h_p\}}{\max \{h_a, h_m, h_p\}},
\end{equation}
$G$ values provide an easy to interpret continuous index, whereas existing methods are usually trained to predict a binary label extracted from radiological reports \citep{tomita-rnn,zebra} or multiclass labels based on threshold levels for $G$ \citep{valentinitsch2019opportunistic,burns2017vertebral}. A related problem is the interpretability of the methods' outputs. The only available information is the network's attention \citep{tomita-rnn} or a similar score \citep{nicolaes-segmentation} somehow related to the probability of fracture presence. At the same time, the medical community is not satisfied with the level of interpretability of such approaches \citep{ghassemi2021false}.

\textbf{Our contribution} is twofold. 
First, our method estimates six keypoints to detect each vertebra and estimate its heights $h_*$ simultaneously (Fig. \ref{fig:method}c-e), which results in excellent fracture classification quality with the area under ROC curve equal to $0.95$ or higher. The predictions are highly interpretable as they can be validated by a doctor using a simple ruler. 
Second, we demonstrate the generalizability of our approach by evaluating the vertebrae detection model on the VerSe \citep{verse1} dataset \textit{without any training/fine-tuning}, which results in only a minor drop of fracture severity classification quality.

This work extends our previous conference paper \citep{neuro-ml-backbone} published at MICCAI-2020. The additional primary contributions are
\begin{itemize}
    \item We propose a new anchor-free approach to detect vertebrae. We compare this idea with our previous anchor-based method and show that it offers not only a more elegant but also more accurate algorithm, see Section \ref{sec:results_cancer50}.
    \item We tested the generalizability of the proposed fracture quantification method in two ways. First, we transfer it from LungCancer-500 to VerSe with a negligible drop in quality. Second, we tested the proposed approach within the mosmed.ai initiative, an experiment for systematic comparison of various AI solutions for medical imaging \citep{pavlov2021reference}.
    \item We extended the previous public release of vertebra annotation on LungCancer-500 dataset by adding missing annotations for lumbar and cervical vertebrae.
\end{itemize}

\section{Previous Work}
\label{sec:previous}

We split previous methods into three major groups:
\begin{itemize}
    \item \textit{Vertebra-level analysis}. The goal of these methods is to estimate severity for every vertebra. Usually, these methods rely on an external vertebra detection method to provide a small 3D path for each vertebra.
    \item \textit{Patient-level analysis}. These methods operate with the whole 3D image and assign a label to the whole CT series. Though vertebra-level predictions can be aggregated, e.g., by max pooling, some of the methods work with a single classification label for the whole 3D image, so it cannot be decomposed into a series of vertebra-level predictions. Patient-level methods usually provide the full pipeline and don't use external vertebra localization as an input.
  \item \textit{Other spine-related methods}. Some interesting ideas are proposed in other spine imaging analysis problems, which aren't directly related to fracture detection.
\end{itemize}

\subsection{Vertebra-level fracture classification}
The automatic classification of vertebral fractures has received much attention from the medical image analysis community. A quantitative image analysis method was proposed in \citep{burns2017vertebral} to classify individual vertebrae. First, the spinal column is segmented by an external method detecting intervertebral discs. Then each vertebra is split into 17 sections to extract a set of simple features such as mean density from the segmentation mask. 
Finally, a support vector machine classifies vertebrae based on the obtained 51 features. The system provides excellent sensitivity (98.7\%) but quite low specificity (77.3\%). 
A similar approach was used in \citep{valentinitsch2019opportunistic} where authors calculated  computer vision features such as histograms of oriented gradients from vertebra masks and achieved ROC AUC 0.88. A plain deep learning-based version of this two-step approach was proposed in \citep{antonio2018vertebra}, where classical ResNet was trained on 3-channel 2D images obtained from the prior segmentation mask by taking central sagittal, axial and coronal slices for each vertebra. 

Finally, in \citep{grading-loss} a severity-aware training procedure is proposed. The authors use a tripled-loss-inspired loss function which motivates the network to cluster the representations of each image/patch with similar severities according to their Genant index. This significantly improves the learnt representations, which is empirically demonstrated by training an SVM classifier for the healthy/fractured task. However, because the solved task is classification at either vertebra or image level, the method still lacks interpretability and ease of clinical validation as other classification-based approaches.

It is important to note that all the methods above rely on prior segmentation, which may result in removing some cases with severe abnormalities. Indeed, the authors of \citep{valentinitsch2019opportunistic} reported that 11 cases out of 154 were excluded from the analysis due to incorrect prior spine segmentation largely caused by high-grade fractures.

\subsection{Patient-level fracture classification}

This requirement was relaxed in several papers. In \citep{nicolaes-segmentation} the authors proposed a two-step pipeline for vertebrae detection: first, a segmentation neural network is used to generate pixel-level predictions (background, normal, fracture), then the predicted maps are aggregated. Instead of the whole spine mask, the authors used the ground-truth coordinates of vertebrae centroids to produce vertebrae-level predictions and achieved ROC AUC 0.93.  A simple idea was used in \citep{tomita-rnn}, where the authors selected the central sagittal slices as the spine is usually located in the middle of the image. In particular, they processed only 6.9 central slices per study (on average). As a result, this approach fails to identify fractures in patients with at least moderate scoliosis, and they had to exclude 156 out of 869 subjects from the analysis, primarily due to scoliosis. Though the average prevalence of scoliosis is 8.85\%, it positively correlated with age and increases from 10.95\% in 60-69 to 50\% in 90+ age groups \citep{kebaish2011scoliosis}, so this cohort can not be ignored in vertebral fractures screening. The classification method from \citep{tomita-rnn} consists of a ResNet34 which processes each of the central sagittal slices separately; then the obtained scores are aggregated by a simple LSTM network. 

An original approach was proposed in \citep{zebra}. Though the method also relies on external spine segmentation, the mask is used to extract the spinal cord and create a new virtual sagittal slice. Next, small patches are extracted from this slice and classified by a convolutional network; finally, a recurrent neural network (RNN) is used to aggregate the predictions from each patch. Although the training database is the largest among the reviewed works (consisting of 1673 cases), the model achieves 83.9\% sensitivity (with 93.8\% specificity), likely due to poor study-level binary annotation extracted from the radiological reports.

This approach was further extended in \citep{seq-to-seq-backbone}. First, a YOLO-like object detection network is used to localize the vertebrae on axial slices; the resulting locations are then linearly interpolated to obtain the spinal cord location. The localized spinal cord is split into multiple volumetric patches to tile the vertebrae with minimal overlap. Next, a patch-wise network is used to obtain a fixed-shape representation for each patch. Finally, an aggregation network maps these representations to a final label (healthy/fractured). The patch-based approach gives the possibility to combat various exclusion criteria such as scoliosis. On the other hand, patch-level predictions cannot be accurately mapped to individual vertebrae, giving only rough localization; thus only a single label per image can be predicted, which greatly reduces interpretability and potential for clinical validation.

Another interesting patient-level method was proposed in \citep{yilmaz2021automated}. It follows a similar high-level scheme and employs a hierarchical convolutional network \citep{buerger2020combining} to localize vertebrae. Then a simple 4-layer CNN analyzes patches to estimate vertebrae deformity and fracture grades. Finally, the obtained vertebra-level scores are aggregated via maximum function. 

\subsection{Other spine-related methods}

An interesting vertebrae detection and labeling idea is proposed in \citep{windsor-backbone}. First, a 2D network is separatelly applied to sagittal slices in order to (1) detect vertebrae corners and centroids and (2) assign corners to their respective centroids: each "corner" pixel is treated as a potential corner of a vertebra’s bounding quadrilateral and 4 displacement fields (for each corner type) are predicted. The displacement fields are then used to group the corners pointing to the same centroid. Finally, the obtained vertebrae are labeled using 2 additional convolutional neural networks combined with a language model, which, by design, guarantees the monotonicity of the resulting labels. However, in 
addition to its complexity, the approach doesn't always yield geometrically valid predictions, because the corners are not originally tied to centroids. For this reason the authors applied additional post-processing in order to alleviate the problem of too many or too few corners per quadrilateral.

Another iterative approach for vertebrae segmentation is discussed in \citep{iterative-backbone-segmentation}: first a 3D segmentation network is used to roughly localize the thoracic, lumbar and cervical vertebrae. Next, an iterative convolutional network is used to segment individual vertebrae in each region as well as predict the rough location of the next vertebra.

A carefully designed pipeline for vertebra identification and labelling was proposed within VerSe-2020 competition \citep{sekuboyina2021verse} by Christian Payer who won the challenge.
First, the spine is localized by a spinal centerline heatmap regression predicted by U-net. Second, SpatialConfiguration-Net \citep{payer2019integrating} is employed to detect centres of the vertebral bodies as landmarks by combining its local appereance with global joint configuration of all vertebrae.

\section{Method}

The high-level structure of the method follows our previous work \citep{neuro-ml-backbone} as we use a 2-stage pipeline. 
\begin{enumerate}
    \item Spine localization. We propose a new soft-argmax based approach to identify the vertebral column in 3D CT and, as a consequence, reducing the problem to 2D by producing the corresponding mid-sagittal slice \citep{buckens2013intra} to measure $h_a, h_m, h_p$ for each vertebra (Fig. \ref{fig:method}a,b). Our method is trained to directly solve the localization problem rather than spine segmentation and demonstrates excellent localization quality with the average error less than $1$ mm. Also, it allows us to process all studies with no exceptions, including cases with severe scoliosis. 
    
    This step is similar to other recent pipelines: the VerSe-2020 \citep{sekuboyina2021verse} winning solution by C. Payer employs heatmap regression to detect spinal center-line; \citep{seq-to-seq-backbone} detects the center-line points by a 2D YOLO. 
    In fact, both soft-argmax and heatmap regression are actively used in landmark detection while the former approach shows better results in some pose estimation tasks \citep{luvizon20182d}. Detection methods like YOLO seems to be less appropriate as we need to find just one point for every slice, not an arbitrary number of objects. 
    \item Vertebra detection and fracture quantification. The second task at hand is very similar to object detection: for a given image it is required to localize all objects of a given class, the number of objects may vary, however it is limited by a constant.

    The main difference is the encoding of sought objects: in classical object detection axis-aligned bounding boxes (AABBs) are used, while in our case each vertebra is represented by 6 coplanar points in an image.
    
    We propose a modified 2D object detection network to predict the Genant segments directly.
\end{enumerate}

Our contribution is the following:
\begin{itemize}
    \item We lift the previous limitation to chest CT images by incorporating the information regarding the \textit{vertebrae limits} into the first network, thus making it applicable to any input.
    \item We simplify the second network by \textit{removing the anchor boxes}.

\end{itemize}

\subsection{Spine localization}
\label{sec:localization}

\begin{figure}
    \begin{center}
      \includegraphics[width=1.\linewidth]{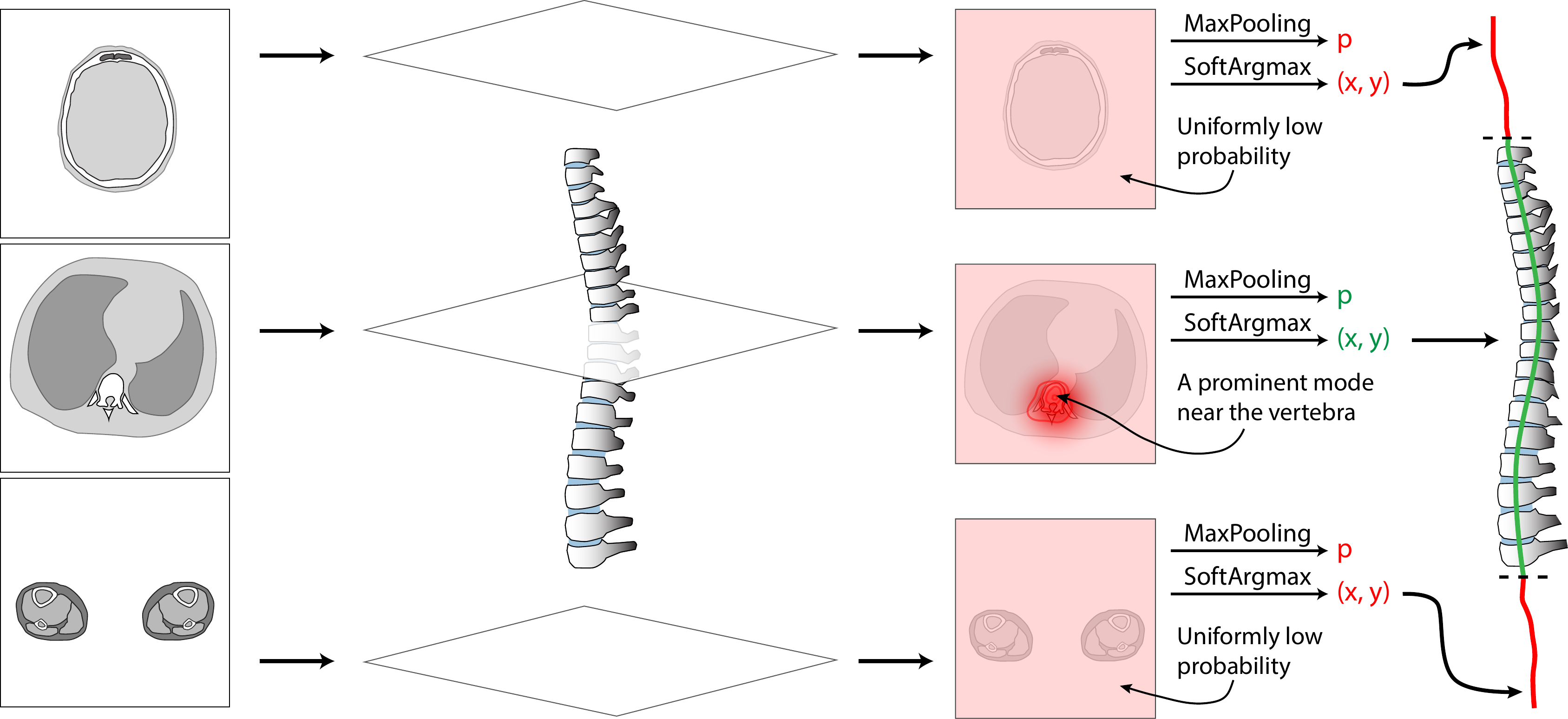}
      \caption{
      The spine localization pipeline:
      a) three axial slices from different body regions: head, thorax, legs; 
      b) the slices' spatial position relative to the spine; 
      c) the predicted probability maps for each slice, the color intensity denotes the probability magnitude;
      d) the final curve passing through the vertebral column (green), as well as the redundant parts (red) delimited by the spine limits (black, dashed).
      }
      \label{fig:localization}
    \end{center}
\end{figure}

We start our pipeline by localizing the spine. For this purpose we use a 3D UNet-like \citep{resunet} fully convolutional neural network. For each voxel we interpret the network's output as the probability of being situated near the vertebral column. Next, we process the prediction in two ways (Fig. \ref{fig:localization}):

\begin{enumerate}
    \item We take the 2D soft-argmax \citep{soft-argmax} operation along the xOy axes in order to obtain the spine coordinates in axial planes.
    \item We take the global max-pooling operation along the same axes in order to obtain the probability of containing the spine at a given $z$ coordinate.
\end{enumerate}


We train the model by optimizing a sum of two loss functions: 

\begin{enumerate}
    \item \textit{Mean absolute error} between the predicted coordinates and the ones smoothly interpolated between vertebrae centers, calculated from annotation (Fig. \ref{fig:target}a).
    \item \textit{Binary cross-entropy} between the predicted slice-level probabilities and the binary limits, also extracted from the vertebrae annotation.
\end{enumerate}

At inference time we threshold the slice-level probabilities by $0.5$ and take the convex hull in order to obtain the curve limits. The predicted curve is then cropped according to these limits. No additional post-processing is used.

\subsection{Spine straightening}
\label{sec:straightening}

\begin{figure}
    \begin{center}
      \includegraphics[width=1.\linewidth]{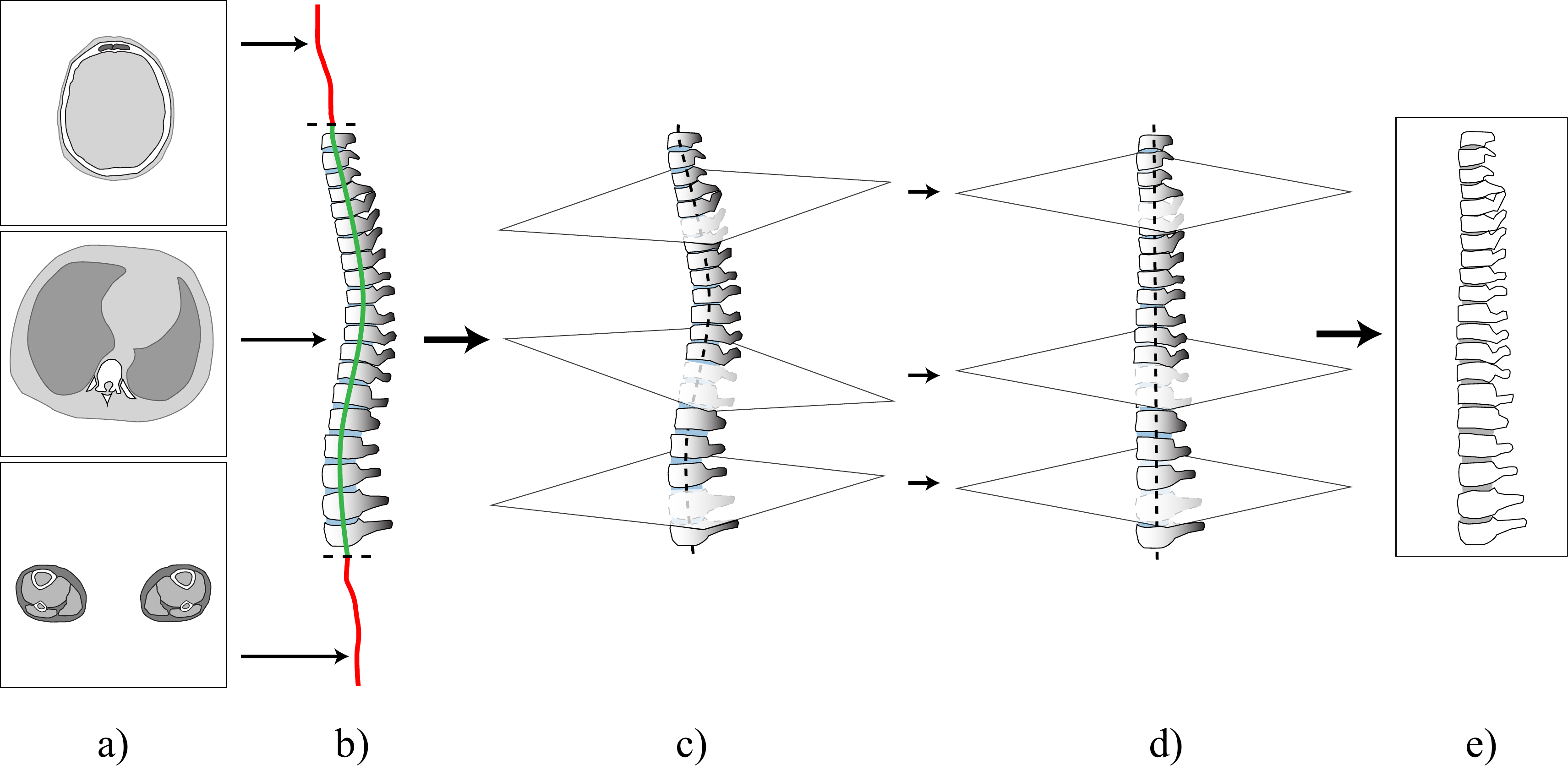}
      \caption{
      The spine straightening pipeline: 
      a) three axial slices from different body regions: head, thorax, legs; 
      b) the combined points from each slice result in a 3D curve, colors denote the probability of being inside the vertebral column limits: green - high probability, red - low; 
      c) planes orthogonal to the curve (for better visualization most planes are omitted);
      d) a straightened vertebral column (the planes become parallel);
      e) the new central sagittal plane.
      }
      \label{fig:straighten}
    \end{center}
\end{figure}


Using the predictions from the first network, we obtain a 3D curve passing through the vertebral column, as well as the limits in which it is defined.
We then crop the image accordingly and interpolate it onto a new 3D grid on which the obtained curve becomes a straight vertical line. 

In order to find such a grid, we select a number of equidistant points on the curve and construct corresponding orthogonal planes. Then, we generate a grid for which all the planes become parallel, which effectively straightens the curve, because the plane normals are tangent to the curve\footnote{See \url{https://github.com/neuro-ml/straighten} for full code for interpolation along curves.}.
Finally we select a new sagittal plane where all vertebrae are visible, namely the one that contains the entire curve. Fig. \ref{fig:straighten} shows a detailed illustration.


\subsection{Vertebrae localization}
\label{sec:detection}

In classical object detection axis-aligned bounding boxes (AABBs) are used as a relatively adequate description of both localization and shape of a given object. 
In our case the Genant segments play the same role while also containing more task-specific information, namely the level of deformation. This fact suggests that AABBs can be completely removed from our training pipeline. 

\begin{figure}
    \begin{center}
      \includegraphics[width=1\linewidth]{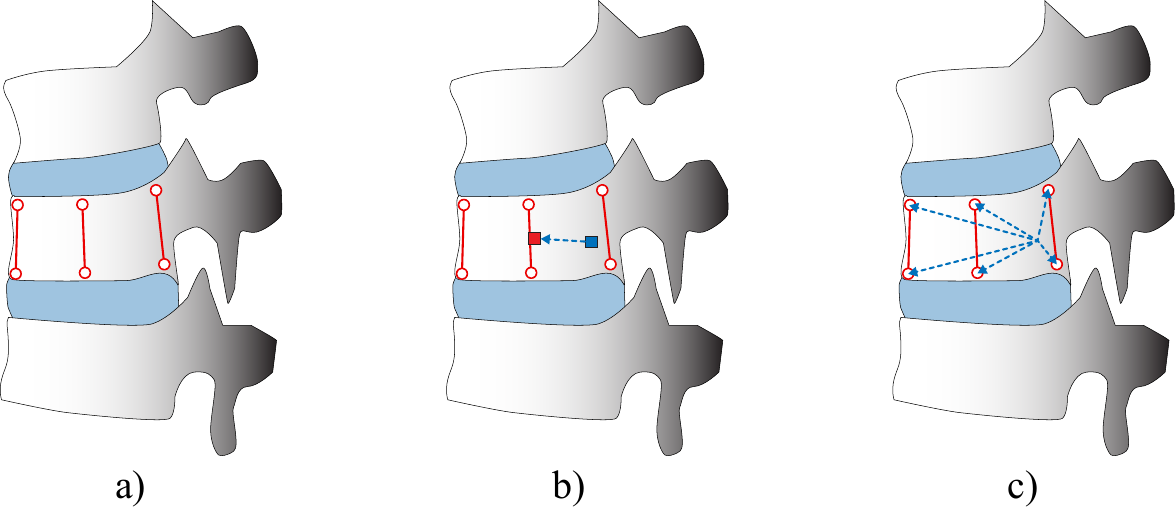}
      \caption{Target generation steps: 
      a) example of an annotated vertebra; 
      b) the distance between the anchor pixel (blue square) and the vertebra centroid (red square); 
      the \textbf{objectness} $O$ is 1, if the distance is smaller than a given threshold; 
      c) the \textbf{keypoints' coordinates} relative to the anchor pixel.}
      \label{fig:target}
    \end{center}
\end{figure}

During \textbf{target generation} we simplify the idea from \citep{faster-rcnn} and use anchor-based translation-invariant encoding (Fig. \ref{fig:target}b-c): each pixel is treated as an anchor relative to which the 6 points' coordinates are calculated:

\begin{equation}
    \label{eq:keypoints}
    e^x = g^x - a^x; \quad e^y = g^y - a^y,   
\end{equation}
where $(g^x, g^y), (e^x, e^y)$ are the global and encoded coordinates of a given point respectively and $(a^x, a^y)$ - are the coordinates of an anchor pixel.

Additionally, as in standard object detection, each location requires an \textit{objectness} label, which is used to filter out the pixels not related to any vertebrae. The objectness is 1 if the given pixel is closer than a fixed threshold to any vertebra centroid (Fig. \ref{fig:target}b), and 0 otherwise.

Finally, we use the same \textbf{loss function} as in \citep{neuro-ml-backbone} to train our second network:
\begin{equation}
    \label{eq:loss}
    L = BCE(\hat o, o) + 
     \frac{1}{\sum I[o_i = 0]}\sum\limits_{i=1}^{N}
      \frac{I[o_i > 0]}{G_i} \cdot MAE(\hat e_i, e_i),
\end{equation}
where 
$BCE$ is the \textit{log-loss} between the real ($o$) and predicted ($\hat o$) objectness,
$MAE$ is the \textit{mean absolute error} between real ($e_i$) and predicted ($\hat e_i$) encoded keypoints' coordinates (\ref{eq:keypoints}) for the i-th vertebra and $G_i$ is the respective Genant score (\ref{eq:genant}) used for loss reweighting. We found such a reweighting of regression loss to be very effective in balancing the network's performance across vertebrae with different fracture severities.


\subsection{Non-maximum suppression}
\label{sec:nms}

\begin{figure}
    \begin{center}
      \includegraphics[width=1.\linewidth]{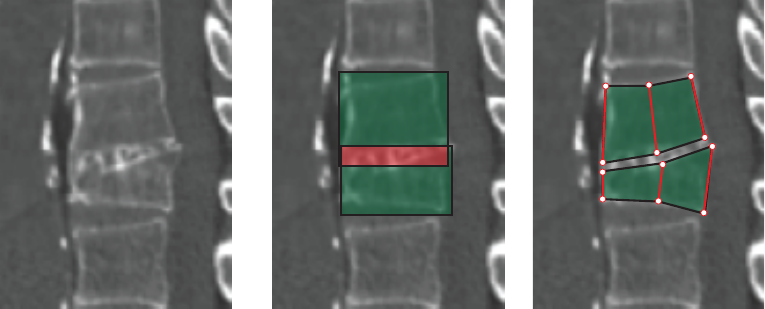}
      \caption{
      A comparison of IoU and BIoU for two deformed vertebrae:
      a) a portion of the vertebral column on a straightened image; 
      b) vertebrae bounding boxes, the red region denotes their intersection; 
      c) vertebrae "butterfly" hexagons, built using the Genant segments (red) note the empty intersection and thus zero BIoU. 
      }
      \label{fig:biou}
    \end{center}
\end{figure}

Non-maximum suppression (NMS) is an essential component of the majority of current object detection pipelines. Its main purpose is to reduce the number of predictions referring to the same object, and ideally leaving exactly one final prediction per object. NMS requires a measure of \textit{closeness} between two given predictions, i.e. a measure of overlap between bounding boxes, as well as an \textit{objectness} score.



In our case, the objectness score is directly predicted by the model.

As for closeness, we replace the intersection-over-union (IoU) between axis-aligned bounding boxes from \citep{neuro-ml-backbone} by new a measure, named as \textit{Butterfly-IoU} (BIoU).
In general, we compute BIoU in the same way as standard IoU, but instead of axis-aligned bounding boxes (Fig. \ref{fig:biou}b), we use hexagons built on six vertebra keypoints. First, we build these hexagons, as shown on Fig. \ref{fig:biou}c, for two sets of key points. Second, we calculate areas of intersection and union of two hexagons. Finally, we divide the intersection area by the union area to calculate BIoU. During non-maximum suppression, we use BIoU values to remove similar predictions as usual.

BIoU is specifically designed to better capture the shape of vertebrae and yields significantly better results in cases of severely damaged vertebrae (Fig. \ref{fig:biou}).

\section{Data}
\subsection{LungCancer500}
\label{dataset:lung-cancer-500}

Our main dataset consists of 100 randomly selected images from the \textit{Moscow Radiology CTLungCa-500} dataset\footnote{https://mosmed.ai/en/data-sets/ct\_lungcancer\_500/} \citep{lung-cancer-500}. The cohort includes studies from a lung cancer screening program, so osteoporotic fractures are common incidental findings as patients aged from 50 to 75.

The images have various voxel spacing ranging from $.5 mm \times .5 mm \times .8 mm$ to $1mm \times 1mm \times .8mm$ and different numbers of visible vertebrae: from 10 to 15.

In \citep{neuro-ml-backbone} an annotation of genant segments for this dataset was published. However, the work was focused solely on detection of thoracic vertebrae. We extended the annotation by adding vertebrae from the remaining regions\footnote{https://github.com/neuro-ml/anchor-free-genant/}.
The re-annotation was performed by 3 experts with $1$ to $5$ years of experience in radiology and a board-certified radiologist with 12 years of experience in the field. In total the dataset contains 3565 vertebra annotations (2-3 per single vertebra). To annotate vertebra heights, radiologists (1) look for a sagittal slice passing through the middle of the vertebra and (2) marks six keypoints on this slice. Then the next vertebra is annotated in the same manner, but the selected slice can be different if a patient has scoliosis or patient positioning is wrong.

The distribution of vertebral fractures is the following: 440 mild, 250 moderate, 54 severe deformations and 2821 normal vertebrae.
Patient-wise we have a somewhat balanced distribution with 11, 23, 44 and 22 patients with none, mild, moderate and severe deformations respectively. 

\subsection{VerSe-2020}
\label{dataset:verse2020}

For additional validation on external data 
we use the \textit{VerSe-2020} dataset \citep{verse1,verse2,verse3,verse4,verse5}. The dataset consists of over 300 multidetector CT images of various regions of the spine.

For each image, every vertebra has an associated segmentation mask as well as centroid coordinates. Additionally, a subset of VerSe-2020 contains vertebrae deformation labels calculated using the same Genant scale. However, the vertebrae near image borders sometimes lack annotation, which mostly impacts the estimated precision of our method (see Section \ref{sec:results} for details). Additionally, because during training we rely on interpolation \textit{between} vertebrae, the image containing a single annotated vertebra (verse116) was excluded from the training set.

\subsection{Private dataset}
In addition to public datasets, we also tested the pipeline trained on a large private dataset which includes 
\begin{itemize}
    \item 402 chest and abdominal CT studies with annotated Genant segments using the same protocol as LungCancer500. The distribution of vertebral fractures is the following: 667 mild, 364 moderate, 153 severe deformations and 4364 normal vertebrae. The annotation was done by three experienced radiologists with at least 10 years of experience. 
    \item 191 additional studies of chest, abdominal and brain CT with annotated limits to improve the spine localization network performance.
\end{itemize}

\subsection{Mosmed.ai test dataset}
In addition to testing the models on public datasets, we also evaluated them on an independent test dataset collected by the Research and Practical Clinical Center for Diagnostics and Telemedicine Technologies of the Moscow Health Care Department.  
The dataset was prepared to test various algorithms using the principles described in \citep{pavlov2021reference} and consists of 50 studies with vertebral fractures and 50 age-matched healthy studies (Genant score is 0.75 or higher). Cases were prepared carefully to test the algorithms under various conditions, such as vertebral ankylosis, vertebroplasty and osteoblastic metastases, among others. 

Only patient-level metrics are available for this dataset. The data is hidden from the developers; the test was conducted in real time with 60 seconds response time requirement. 






\section{Experimental setup}



We trained our \textbf{spine localization} network with Adam optimizer \citep{adam} with standard parameters for $10^5$ iterations with batches of size 3 and a learning rate of $1 \cdot 10^{-4}$ ($3 \cdot 10^{-5}$ for VerSe). As a preprocessing step we normalize the voxel intensities to the interval $[0, 1]$ as well as resample the images to a spatial size of $2 mm\times 2mm \times 4 mm$. Our motivation behind this is to standardize the spatial features, because CNNs are not invariant to scaling, as well as reduce memory consumption for images with too high resolution. 
No additional postprocessing is applied to final predictions, aside from the probability maps binarization described in Section \ref{sec:localization}.


The \textbf{vertebrae detection} network was trained for 80k iterations with batches of size 8. We used Adam optimizer \citep{adam} with standard parameters and a gradually decreasing learning rate, which enabled the models to reach better optima. The initial learning rate was set to $3\cdot10^{-4}$ and decreased by a factor of 2 after 6k, 10k, 16k, 28k, 40k and 56k iterations.
As a preprocessing step we normalize the voxel intensities to zero mean and unit variance. As described in Section \ref{sec:detection}, we generate a grid onto which the new image is interpolated. The grid is constructed in such a way, so that the spatial size of each voxel becomes $1mm\times 1mm\times 1 mm$, which can be regarded as another implicit preprocessing step. Finally, during non-maximum suppression we leave the predictions with an \textit{objectness} greater than 0.7 and use a threshold of 0.1 for the closeness function (\ref{sec:nms}).

\newcommand{\lt}{\textless}

\section{Results}
\label{sec:results}

We report results obtained using 5-fold cross-validation. For every network we trained 5 experiments with different cross-validation splits in order to estimate mean and standard deviation for each score. 
To obtain patient-level predictions, we use the most severe fracture among all the vertebrae, which is equivalent to taking the minimal Genant score.
As we have multiple annotations per study, we also report the inter-expert variability. 

Unlike Lung-Cancer-500, VerSe has a publicly available division on train, validation and test. For this reason, we used the last two subsets for testing. 

\subsection{Spine localization}
\label{sec:spine-localization-results}
We trained models on different datasets with various characteristics:
\begin{enumerate}
    \item Lung-Cancer-500, which consists of pretty standard Chest CT scans primarily with thoracic vertebrae. 
    \item More informative Private and VerSe datasets. As Lung-Cancer-500 contains a very limited number of lumbar and cervical vertebrae, we didn't transfer models trained on Lung-Cancer-500 to VerSe.
\end{enumerate}

We report the localization quality of the first step of our method in Table \ref{tab:spine-localization-metrics}. 
Because most of the images from Lung-Cancer-500 (Section \ref{dataset:lung-cancer-500}) are fully covered by vertebrae, this dataset is not very challenging for the limits classification head.
Thus, in order to thoroughly evaluate the localization network, we use an additional dataset with annotated vertebrae centroids - VerSe-2020 (Section \ref{dataset:verse2020}).

\begin{table}[t!]
\caption{
    \label{tab:spine-localization-metrics}
    Spinal line localization metrics for Lung-Cancer-500 and VerSe-2020 datasets.
}
\begin{center}
\begin{tabular}{cccc}
    \toprule 
   \multicolumn{2}{c}{Data} & \multirow{2}{*}{\specialcell{Points \\ Mean l2, mm}} & \multirow{2}{*}{\specialcell{Limits \\ MAE, mm}} 
   \\ 
    
    \arrayrulecolor{black!30}\cmidrule(lr){1-2}
    \arrayrulecolor{black}
    
    Train&Test && \\ 
    \cmidrule(lr){1-1}
    \cmidrule(lr){2-2} \cmidrule(lr){3-3} \cmidrule(lr){4-4} 
    
    Cancer500 & \multirow{2}{*}{Cancer500} &
        \enskip .92 (.07) &
        \enskip .03 (.04) 
        \\
        
    Private &  &
        \enskip .74 (.06) &
        \enskip .07 (.08) 
        \\
        
    \arrayrulecolor{black!30}\cmidrule(lr){1-4}
    \arrayrulecolor{black}

    VerSe & \multirow{2}{*}{VerSe val} & 
        \enskip 1.81 (.20) &
        \enskip 19.23 (3.42) 
        \\
        
    Private &  & 
        \enskip 1.30 (.15) &
        \enskip 18.37 (3.82) 
        \\
    
    \arrayrulecolor{black!30}\cmidrule(lr){1-4}
    \arrayrulecolor{black}

    VerSe & \multirow{2}{*}{VerSe test} & 
        \enskip 1.72 (.17) &
        \enskip 18.76 (2.59) 
        \\
        
    Private &  & 
        \enskip 1.19 (.16) &
        \enskip 18.07 (3.10) 
        \\

    \bottomrule
\end{tabular}
\end{center}
\end{table}

The difference in the Mean l2 metric between the two datasets can be 
explained by the fact that the images of VerSe are much more diverse than those of Lung-Cancer-500. 
Nevertheless, given the input voxel size of $2mm\times 2mm\times 4mm$, the results on both the datasets suggest that the model's performance is close to maximal.


\begin{figure} 
    \begin{center}
      \includegraphics[width=0.8\textwidth]{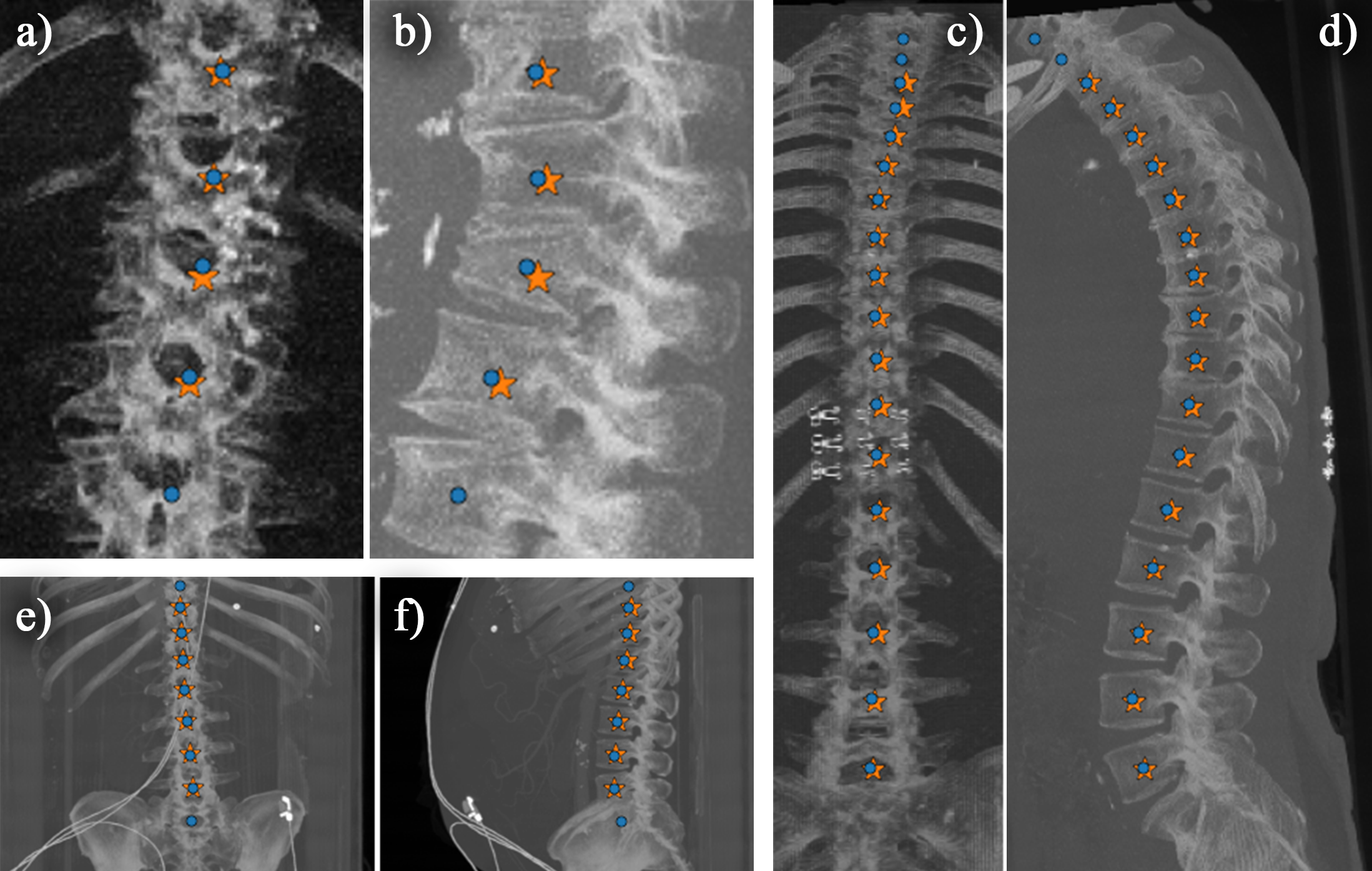}
      \caption{
        Several examples of "false positives" on the VerSe dataset in sagittal and coronal projections. The annotated vertebrae are marked with orange stars, the predicted ones - with blue circles. Note the missing annotations for "border" vertebrae: near image borders (a-b), vertebral column borders (c-f).
        \label{fig:verse}
      }
    \end{center}
\end{figure}

Moreover, as mentioned in Section \ref{dataset:verse2020}, the vertebrae near image limits are sometimes lacking annotation in VerSe, see several examples on Fig. \ref{fig:verse}. We argue that this is the main cause of such a large \textit{Limits MAE} in Table \ref{tab:spine-localization-metrics}. For this reason, we additionally extrapolate the predictions of our spine localization network trained on VerSe by 2cm at both limits, leaving the second network with more potential vertebrae to detect. Surprisingly, this simple post-processing increased the overall recall of the pipeline, but doesn't deteriorate the precision significantly (Table \ref{tab:detection-metrics}).

\subsection{Vertebrae detection and severity classification}

\subsubsection{Cross-validation on Cancer500}
\label{sec:results_cancer50}
We first compare three modifications of the pipeline
\begin{enumerate}
    \item \textit{Anchor-Boxes}. The approach from our previous work \citep{neuro-ml-backbone}. Both spine localization and vertebra analysis networks were trained on Lung-Cancer-500.
    \item \textit{Ours}. The proposed anchor-free approach; both networks were trained on Lung-Cancer-500.
    \item \textit{Ours Private}. The proposed approach; both networks were trained on the Private dataset.
\end{enumerate}

In order to calculate vertebrae-level metrics a matching procedure of predicted and real vertebrae is required. For Cancer500 we use the Butterfly-IoU (namely BIoU) between each $(prediction, target)$ pair, see Fig. \ref{fig:biou} above. 
All pairs with a closeness smaller than a given threshold are discarded. For the remaining pairs, for each $target$ we select the closest corresponding $prediction$. This way all unmatched targets are considered false negatives (FN) and all unmatched predictions - false positives (FP). Table \ref{tab:detection-metrics} shows vertebrae detection metrics averaged by patients as well as by vertebrae.

\begin{figure}[b]
    \begin{center}
      \includegraphics[width=1.\linewidth]{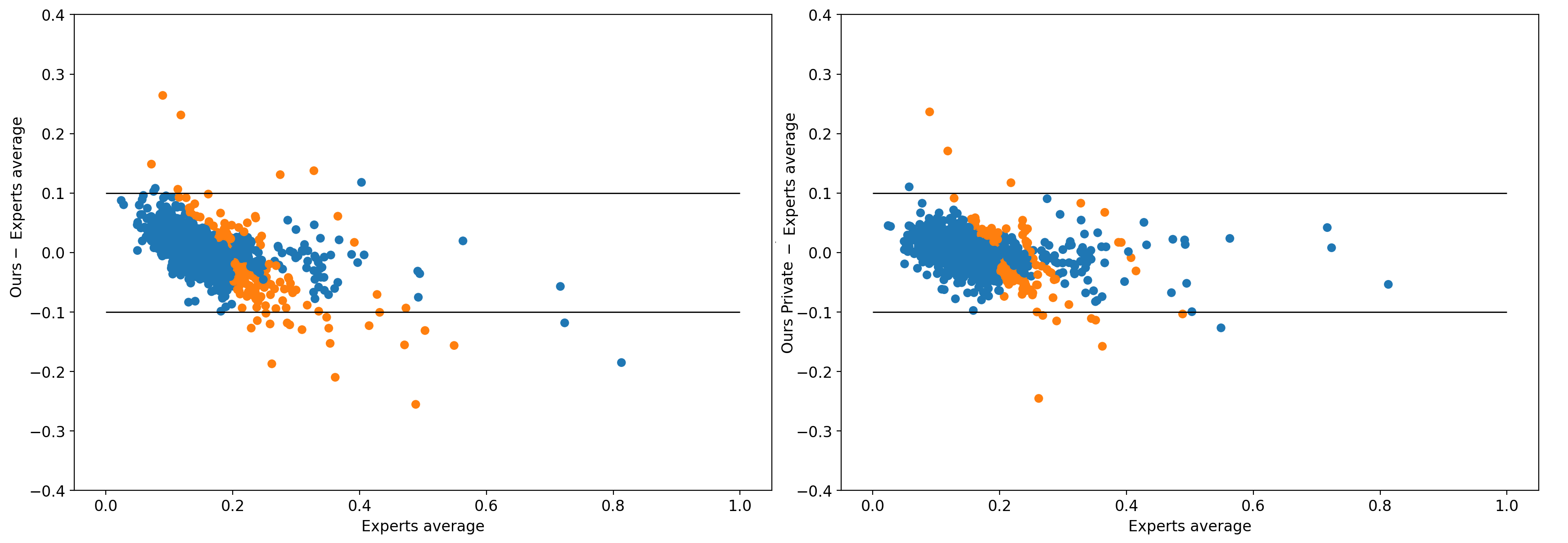}
      \caption{
        Bland-Altman plots for Ours and Ours-Private models. Blue points denotes vertebra with correct Genant grade classification; orange ones shows wrong classification.
        \label{fig:bland-altman}
      }
    \end{center}
\end{figure}

\begin{table}[bt!]
\caption{
    \label{tab:detection-metrics}
    Vertebrae detection and severity classification metrics. \textit{Anchor-Boxes} and \textit{Ours} metrics are based on 5-fold cross-validation. Severity classification metrics are reported on the vertebra level for identifying Moderate and Severe vertebrae ($G\le0.74$). We report sensitivity for a specificity fixed at 90\% to make models directly comparable.
}
\begin{center}
\begin{tabular}{ccccccc}
    \toprule 
    && \multicolumn{2}{c}{Vertebra Detection} & \multicolumn{2}{c}{Severity Classification} \\
    \cmidrule(lr){3-4} \cmidrule(lr){5-6}
   && Precision & Recall & ROC AUC & \specialcell{Sens. at \\spec.=0.9}\\ 
    \cmidrule(lr){2-2} \cmidrule(lr){3-3} \cmidrule(lr){4-4} \cmidrule(lr){5-5}
    
    \parbox[t]{2mm}{\multirow{4}{*}{\rotatebox[origin=c]{90}{\enskip Cancer500}}}&
    Anchor-Boxes &
        \enskip .994 (.001) & 
        \enskip .953 (.001) &
        .955 (.004)&
        .863 (.030)&
        \\

    &Ours & 
        \enskip .991 (.002) &
        \enskip .990 (.002)  &
        .959 (.002)&
        .885 (.002)&
        \\
        
    &Experts & 
        \enskip .999 (.001) &
        \enskip .994 (.001) &
        .971 (.005)&
        .936 (.018)&
        \\

    & {Ours Private} & 
        \enskip .993  &
        \enskip .991  &
        .981 &
        .950 &
        \\

    \arrayrulecolor{black!30}
    \cmidrule(lr){1-6}

    \arrayrulecolor{black}
    
    \parbox[t]{2mm}{\multirow{3}{*}{\rotatebox[origin=c]{90}{VerSe}}}&
     Ours & 
        \enskip .947  &
        \enskip .886   &
        \multirow{2}{*}{.951}&
        \multirow{2}{*}{.848}&
        \\
        
    & Ours ext. & 
        \enskip .935 &
        \enskip .951 &
        &
        &
        \\

    & {Ours Private} & 
        \enskip .896 &
        \enskip .973 &
        .970 &
        .908 &
        \\


    \bottomrule
\end{tabular}
\end{center}
\end{table}

\begin{figure}
    \begin{center}
      \includegraphics[width=1.\linewidth]{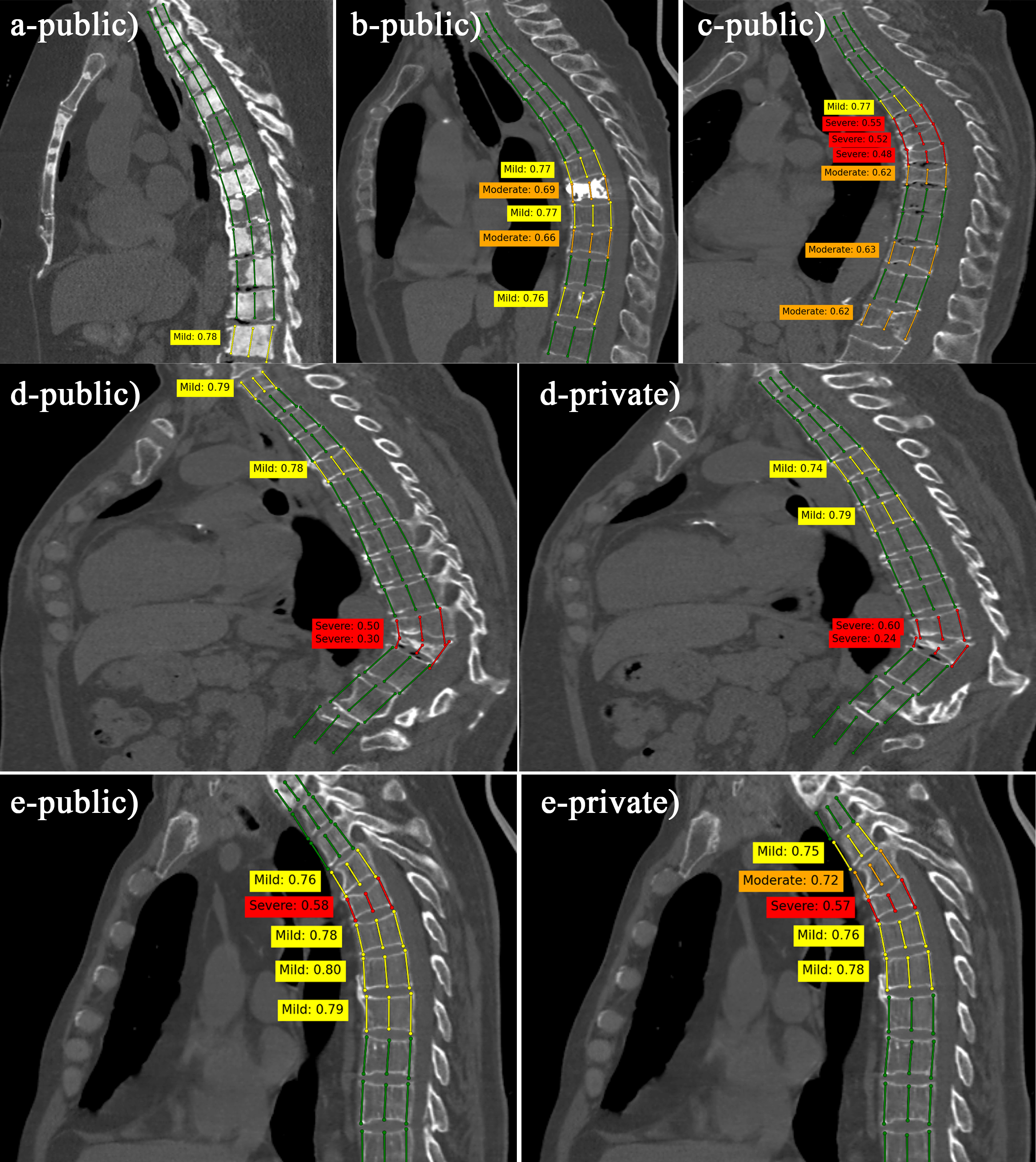}
      \caption{
        Qualitative analysis of predictions for the Cancer500 dataset for both our public and private networks.
        \label{fig:cancer500-analysis}
      }
    \end{center}
\end{figure}

We report two types of metrics: precision and recall for vertebrae identification and binary classification metrics for vertebra fracture severity classification; see the first four lines of Table \ref{tab:detection-metrics}. 
Anchor-free model benefits from higher detection recall and slightly better classification metrics. The same model trained on the Private dataset shows perfect severity classification metrics, even outperforming expert agreement level. Figure \ref{fig:bland-altman} provides some insights about models' performance. Ours-Private models achieves more symmetrical distribution of errors and a generally narrower interval.  Figure \ref{fig:cancer500-analysis} shows some qualitative analysis of predictions for the Cancer500 dataset.


\subsubsection{External tests}
\label{sec:results_external}

For external test on VerSe and mosmed.ai we use two setups:
\begin{enumerate}
    \item \textit{Ours}. Spine localization network is now trained on VerSe as LungCancer-500 contains primarily thoracic vertebrae. The second network is simply reused without any adaptation to VerSe.
    \item \textit{Ours-Private} is used without any modifications as its training data set is diverse enough.
\end{enumerate}

For VerSe another matching procedure is required, given the fact that the annotation represents vertebrae centers. Similarly to \citep{verse1} we use the Euclidean distance between centers as a closeness function with a threshold of 20 mm.

In addition to vertebra-level, we also report patient level by using maximal deformation $1-G$ to determine the whole image label.
To analyze the performance of vertebrae fracture severity classification, we report metrics for various threshold values of $G$ following the radiological definition of severity \citep{genant}, see Table \ref{tab:severity-metrics}. We assume that the most relevant problem for chest CT is the identification of at least Moderate fractures ($G\le0.74$) as healthy vertebrae in the thoracic spine are wedged, so normal variation can be misclassified as a Mild fracture ($0.74<G\le0.8$) \citep{lenchik2004diagnosis}. To enable direct comparison with \citep{grading-loss} we also add \textit{G0 vs G2-G3} where these Mild fractures are removed from the data.

\begin{table}
\caption{
    \label{tab:severity-metrics}
    Binary classification metrics on VerSe for various grades of fractures: at least \textit{Mild} ($G \le 0.8$) and at least \textit{Moderate} ($G \le 0.74)$.
    All numbers are given as mean (std).
}
\begin{center}
\setlength\tabcolsep{0.4em}
\begin{tabular}{cccccc}
    \toprule 


    
    &&\multicolumn{2}{c}{Vertebra-level} &\multicolumn{2}{c}{Patient-level}\\
    \cmidrule(lr){1-2} \cmidrule(lr){3-4} \cmidrule(lr){5-6} 

    Task & Model & ROC AUC & \specialcell{Sens. at \\spec.=0.9} & 
              ROC AUC & \specialcell{Sens. at \\spec.=0.9} \\ 
    
    \arrayrulecolor{black!30}\cmidrule(lr){1-6}
    \arrayrulecolor{black}
    
    \multirow{2}{*}{\specialcell{G0 vs G1,\\ G2, G3}}
    & Public &
        .877 & 
        .713 & 
        .882 & 
        .703 \\ 
        
    & Private &
        .906 & 
        .777 & 
        .911 & 
        .807 \\ 
        
    \arrayrulecolor{black!30}\cmidrule(lr){1-6}
    \arrayrulecolor{black}
    
    \multirow{2}{*}{\specialcell{G0 vs \\ G2, G3}}
    & Public &
        .963 & 
        .889 & 
        .953 & 
        .856 \\ 
        
    & Private &
        .979 & 
        .952 & 
        .962 & 
        .919 \\ 
        
    \arrayrulecolor{black!30}\cmidrule(lr){1-6}
    \arrayrulecolor{black}
    
    \multirow{2}{*}{\specialcell{G0, G1 vs \\ G2, G3}}
    & Public &
        .951 & 
        .848 & 
        .936 & 
        .806 \\ 
        
    & Private &
        .970 & 
        .908 & 
        .960 & 
        .910 \\ 
    
    \cmidrule(lr){1-6}
    
    mosmed.ai & Private &
        N/A & 
        N/A & 
        .99 & 
        1.0 \\ 

    \bottomrule
\end{tabular}
\end{center}
\end{table}

Moreover, we analyze the generalizability of our approach by evaluating the model on the VerSe dataset, which contains additional annotations regarding the fracture severity of each vertebra. In this experiment we only trained the spine localization network on VerSe and reused the vertebrae detection network (trained on Cancer500) \textit{as is}, without any additional tuning. The results are shown in the last rows of Tables \ref{tab:detection-metrics} and \ref{tab:severity-metrics}. Note the minimal drop in classification performance.

It is important to note that the results for the VerSe dataset from Table \ref{tab:detection-metrics} are underestimated, especially the precision. This significant drop in quality is due to lacking annotations for vertebrae near images' edges. Fig. \ref{fig:verse} shows several examples with such ``false positives''. According to our calculations about $95\%$ of FPs are due to such partially annotated cases.

Similar values of ROC AUC were obtained at vertebra ($0.88$ \citep{valentinitsch2019opportunistic}, $0.93$ \citep{nicolaes-segmentation}) and patient levels ($0.92$ \citep{tomita-rnn}). 

\begin{figure}
    \begin{center}
      \includegraphics[width=1.\linewidth]{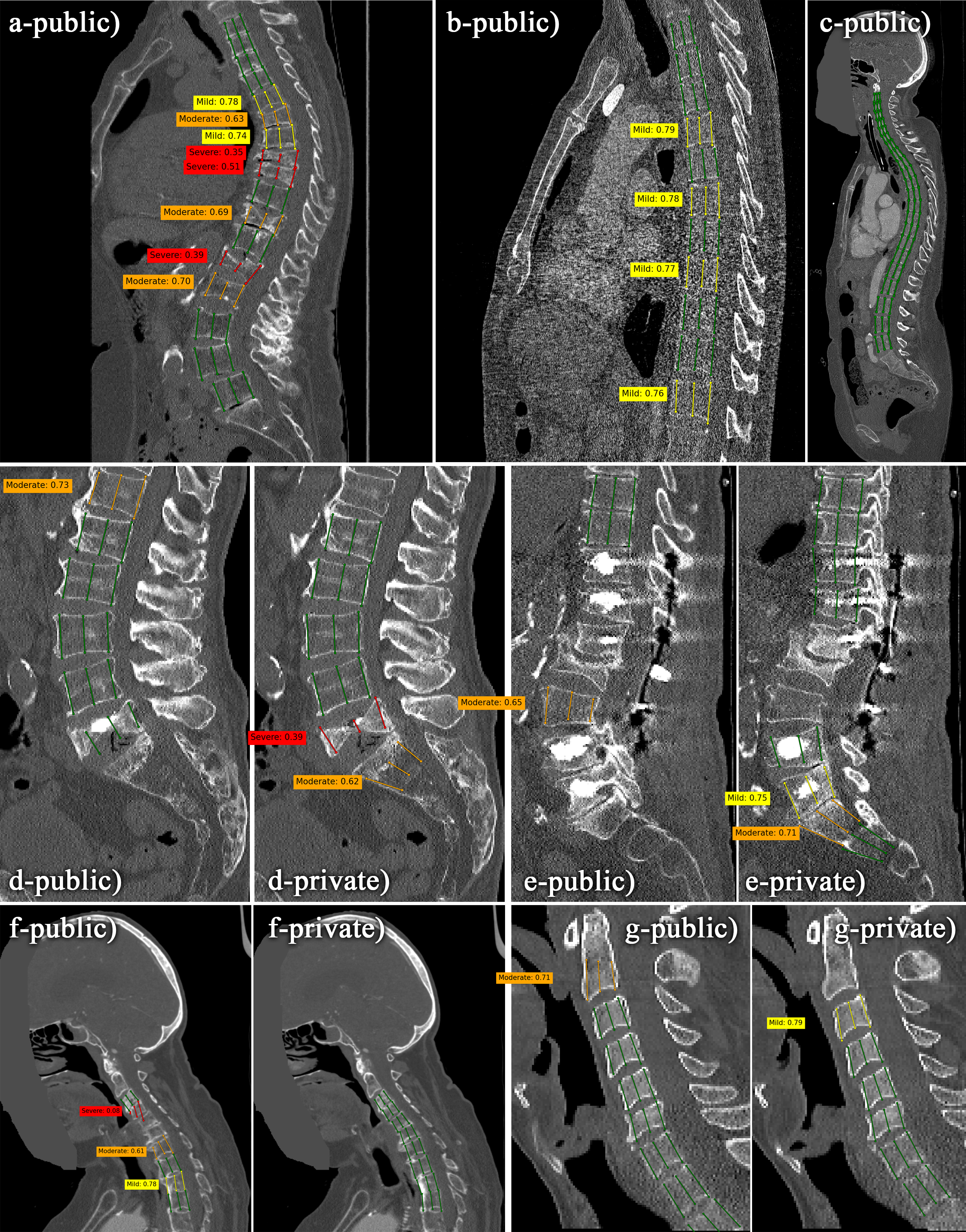}
      \caption{
        Qualitative analysis of predictions for the VerSe dataset for both our public and private networks.
        \label{fig:verse-analysis}
      }
    \end{center}
\end{figure}

Finally, Fig. \ref{fig:verse-analysis} gives a qualitative analysis of our method's performance on VerSe by showing several interesting (both bad and good) predictions. The rest of predictions on the entire VerSe test subset can be found in a separate repository\footnote{https://github.com/neuro-ml/anchor-free-genant/}.

We can see that the network is robust to multiple severe deformations (Fig.~\ref{fig:verse-analysis}a), extreme noise (Fig.~\ref{fig:verse-analysis}b), generalizes well to whole body scans, various regions of the vertebral column and presence of contrast (Fig. \ref{fig:verse-analysis}c).

On the other hand, some predictions suffer from severely misplaced keypoints, mainly in the cervical and lumbar regions. An interesting example is Fig. \ref{fig:verse-analysis}g, which shows a faulty prediction for the C2 vertebra, which has an atypical shape as compared to other vertebrae. Another good example is Fig. \ref{fig:verse-analysis}e, which has multiple false-negatives due to a noticeable amount of artifacts. Also, Fig. \ref{fig:verse-analysis}f shows a case with multiple problems, all of which were caused by the spine localization network.

Overall, our analysis shows that, at a great extent, most of the causes of bad performance can be alleviated by training the network on a larger and more challenging dataset. See Fig. \ref{fig:verse-analysis}d-g for a comparison between our public and private networks.

\section{Discussion}

We proposed an interpretable method for vertebral compression fractures quantification. It's based on clinical recommendations and provides easy-to-verify outputs. 

We extended and simplified the method for automatic identification of vertebrae-level fractures classification using the Genant score proposed in \citep{neuro-ml-backbone}. First of all we demonstrated, for the task at hand, the redundancy of bounding-boxes, and showed that Genant segments represent a more suitable description of vertebrae shape and localization. Furthermore, we extended the spine localization model by adding limits prediction, which makes it more suitable for clinical applications.

Finally, we demonstrated, the generalizability of our approach by evaluating our vertebrae detection network on the publicly available dataset VerSe-2020.



\paragraph{Acknowledgments} Alexey Zakharov, Maxim Pisov, Victor Gombolevskiy and Mikhail Belyaev were supported by the Russian Science Foundation grant 20-71-10134. 

\bibliography{main}

\end{document}